\newcommand{\bicho}{GRS~1758--258}
\documentclass{aa}  

\usepackage{graphicx}
\usepackage{txfonts}
\usepackage{gensymb}
\begin{document}

   \title{Effects of precession versus instabilities on the jets of \bicho}
\titlerunning{Precession vs instabilities in \bicho}
   \subtitle{}

   \author{Pedro L. Luque-Escamilla\inst{1},
           Josep Marti \inst{1}
          \and
           Jos\'e Mart\'inez-Aroza
          \inst{2}\fnmsep
%          \thanks{Just to show the usage
%          of the elements in the author field}
          }
\authorrunning{Luque-Escamilla, Martí \& Mart\'inez-Aroza}
   \institute{FAEG,
             EPS Ja\'en. University of Ja\'en (Spain).\\
              \email{peter@ujaen.es},\\
              \email{jmarti@ujaen.es}
         \and
             FAEG, Facultad de Ciencias. University of Granada (Spain).\\
              \email{jmaroza@ugr.es}
 %            \thanks{The university of heaven temporarily does not
 %                    accept e-mails}
             }

   \date{Received August 25, 2020; accepted March 16, 2021}

% \abstract{}{}{}{}{} 
% 5 {} token are mandatory
 
  \abstract
   {}
   {The prototypical microquasar \bicho\ exhibits large-scale morphological changes in radio maps over time which have been attributed to the rise of instabilities. Here, we investigate whether these effects could be attributed to jet precession instead. }
   {We used new and archival radio maps to fit a kinematic jet precession model. The value of the parameters thus obtained were analysed in order to get constraints on the physical properties of the \bicho\ system. Their consistency with different theories of the origins for the jet precession,  such as Lense-Thirring effect and tidal torques induced by the secondary star, has previously been studied. Alternatively, we also assessed the possibility that observations are compatible with eventual jet instabilities.}
   {The new radio data presented here confirm that the large-scale radio morphology of \bicho\ is changing over time. Our study shows that the 18.48 day period could plausibly be ascribed to precession, thus implying a reinterpretation of  assumptions made for the orbital period to date.  However, the observed structural changes in radio jets cannot be easily attributed to jet precession according to our analysis. In contrast, the growth of instabilities certainly appears to be a more realistic explanation of these effects.}
   {}


   \maketitle
%
%________________________________________________________________

\section{Introduction}

\bicho\ is one of the brightest objects seen in hard X-rays in the close vicinity of the Galactic centre. It was first detected by GRANAT/\textit{SIGMA}  \citep{1991A&A...247L..29S} and since then this system  has enjoyed intense observational coverage at these frequencies over decades from CGRO, RXTE, ROSAT, INTEGRAL, ASCA, Swift, Beppo-SAX, EXOSAT, Chandra, and XMM-Newton \citep[i.e.][]{2000Lin,1994Meregetti,1538-4357-578-2-L129,1997Mereghetti,1999Main,2001Keck,2002Heindl,2006Pott,2001Gold,2011MNRAS.415..410S}. The source appears to exhibit continuous, variable activity at these wavelengths, spending almost all its time in a hard state. The compact object in \bicho\ is probably a black hole as its X-ray spectrum resembles that of the black hole in Cygnus X-1 \citep{1999Main}. All these characteristics have led to consider \bicho\ as a low mass X-ray binary system (LMXB) with a black hole accreting from a stellar companion. Observations in radio wavelengths show two clearly extended lobes, which suggest that \bicho\ is a prototypical microquasar \citep{1992ApJ...401L..15R, 1992Natur.358..215M}. However, the Galactic nature of this system has remained elusive with regard to spectroscopic analysis because of crowded stellar fields, astrometric uncertainties, and significant interstellar extinction towards the Galactic Center in optical and infrared wavelengths \citep{1991Bignami,1992A&A...259..205M,1998Marti,2001Eikenberry,2002Rothstein}.  

Although a weak stellar counterpart of \bicho\ was finally identified \citep{2010A&A...519A..15M,2014ApJ...797L...1L}, it was not possible to obtain any clear spectral signature even using the $10$ m \textit{Gran Telescopio Canarias} \citep{2010A&A...519A..15M, 2016A&A...596A..46M}. Nevertheless, the system was definitively associated to the Milky Way when large-scale morphological changes were noticed by \citet{2015A&A...578L..11M}. %\citet{2014ApJ...797L...1L}. 
These variations in radio shape were attributed to instabilities in the jet and interactions with the interstellar medium, conforming a cocoon observed for the first time in a microquasar, which adds a new aspect of similarity with radio galaxies. A deep inspection of the resulting radio S-shape of \bicho\ revealed it to be a downsized mimic of winged extragalactic radio galaxies, which appear to have some consequences for the black hole merger rate and its corresponding gravitational wave production \citep{2018NosotrosGRS}. 

However, instabilities are not the only mechanisms that could explain the large-scale morphology of astrophysical jets. Precession is often invoked to account for the apparent shape of jets both in extragalactic and Galactic sources. In fact, other known microquasars present observational evidence of jet precession, such as SS\,433 \citep{1979Abell,1981ApJ...246L.141H}, GRO J1655$-$40 \citep{HR95}, and 1E\,1740.7--2942 \citep{2015Nosotros1E}, where the evolution of the large-scale morphology of the jet-lobes is evident over time. Since this mechanism seems to be expected in binary systems similar to \bicho, due to the torque of the donor star \citep{1998MNRAS.299L..32L} or the relativistic drag from the spinning compact object through Lens-Thirring effect \citep{1975Bardeen}, in this paper, we study the possibility that it could be responsible for the changing large-scale radio morphology appearing in \bicho\ at different epochs. 
%In Section 2 we summarise the main properties known at present for this source. 
In Section 2, we show the data and methods we used for fitting the radio jet-lobe changing structures of \bicho\ to a kinematic model, following a similar approach to the one we successfully applied to the case of 1E\,1740.7--2942 \citep{2015Nosotros1E}. We present a complete set of parameters for the jet precession in Section 3 and we analyse the consistency of the results and discuss their consequences in Sections 4 and 5. We present our conclusions in Section 6.

\section{Radio data and precession model fitting}

\begin{figure*}
%   \resizebox{\hsize}{!}
\centering
    \includegraphics[width = \textwidth]{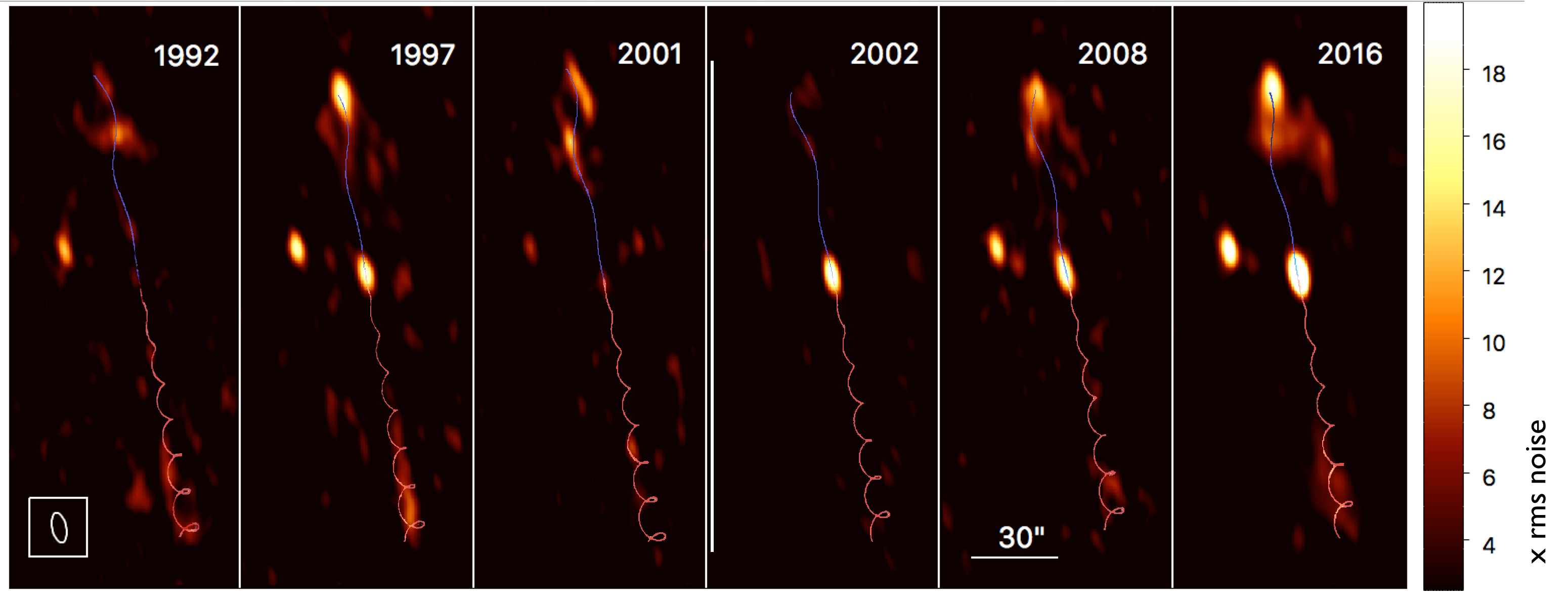}
\hspace{1.6cm}
      \caption{Attempt to fit the large-scale radio jets of \bicho\ jets at different epochs using the theoretical kinematic model of \citet{1981ApJ...246L.141H}. Horizontal bar shows the scale size and the synthesised beam is on the left-down corner. The vertical scale is expressed in units of the root mean squared noise of each radio map that amounts to 9.3, 8.2, 10.7, 16.0, 7.3, and 4.5 $\mu$Jy beam$^{-1}$ for the different epochs, respectively. North is up and east is left. 
              }
         \label{chorros}
   \end{figure*}

As our intention is to study the evolution of the radio jet morphology of \bicho, we reused historical records of this source at radio wavelengths from the public archives of the VLA interferometer. This task had already been carried out for an earlier study of the source, where four maps exhibiting clear morphological variations in time were obtained \citep[see log of observations and detailed calibration methodology in ][]{2015A&A...578L..11M}. In this work, we reuse these four `frames' together with a new one that we obtained from a dedicated campaign on \bicho\ with the VLA in C-configuration conducted in 2016 at the 6 cm wavelength (Project ID. 16A-005, on-source time 7659 s). It is remarkable to note the improvement in sensitivity with regard to this latter observation when compared to the previous ones, which was made possible thanks to the GHz bandwidth available since 2011 following the VLA upgrade, although calibration was applied, taking care to produce a map that would be comparable with previous maps. The final five maps are shown in Fig. \ref{chorros}.  As we can see, the new radio map confirms the previously noticed changes in the large-scale morphology of relativistic jets in \bicho\ \citep{2015A&A...578L..11M}. 

However, the possibility that these variations are due to precession has never been  raised before in the literature. In considering such a case, we tried to fit the well-known precession kinematic model of \citet{1981ApJ...246L.141H} to our data. We follow a similar approach to that successfully used for the case of 1E\,1740.7--2942 \citep{2015Nosotros1E}, thus achieving a multi-epoch fit of a relativistic jet from their projected paths for the second time. 
A detailed description of the fit procedure can be found in this last reference; even though for 1E\,1740.7--2942, we deal with a more continuous radio jets and their location could be well established thanks to the so-called 'skeleton' of the binary image of the maps \citep[see][for details]{2015Nosotros1E}. Unfortunately, \bicho\ presents less continuous jets, so the position of their projection on the maps in Fig. \ref{chorros} need to be determined in a different way. For this purpose, we slice the image and look for maxima in the emission profiles. The jet points thus obtained conform analogous to the so-called jet ridge in active galaxy nucleus (AGN). It is worth mentioning that these points correspond to the traces of radio jet in the maps, not only to the hotspots, which do not move ballistically.

Once the jet locations are known, we
try to fit these data to the kinematic model previously cited. 
It is worth mentioning that all jet paths on the five different epochs have to be simultaneously fit to the same set of model parameters. There are eight different parameters in the kinematic model, so the least-squares fitting procedure is a very time-consuming task. To 
achieve a final result in a reasonable amount of time, we try to narrow the limits on some of the parameters whenever some additional, reliable data is available. 
For instance, the position angle of the jet axis is easily obtained from the maps with enough accuracy to be around $11^{\circ}$. The distance is known to be less than $12$ kpc \citep{2015A&A...578L..11M}, so we let this parameter to vary only up to this value. The inclination $i$ of the jet precession axis with the line of sight cannot be edge-on due to the lack of eclipses in X-ray data, nor completely face-on, because two jets are visible. In order to constrain an inclination value, we can make use of the brightness ratio between approaching and receding continuous, relativistic jets 
defined as $BR = [(1+\beta \cos(i))/(1-\beta \cos(i))]^{2-\alpha}$. Here, $\alpha$ being the spectral index and $\beta$ the ratio $v_{jet}/c$, which, in turn, may be related to the apparent jet velocity as $\beta = \beta_a (\sin(i) + \beta_a \cos(i))^{-1}$. From our deep map obtained by combining all available data at $6$ cm, we measure jet brightness at equal angular separation on both sides to estimate the observed $BR$ to be of about $2.5$, while $\alpha$ ranges from $\sim -0.7$ to $\sim -0.3$ \citep{2002A&A...386..571M}. Apparent velocity is at least $\beta_a = 0.32$ at the end of the jet \citep{2015A&A...578L..11M}. Therefore, the range of possible $\beta \sim 0.32$--$0.33$ while $i \sim 51$\degree --$56$\degree. These inclination angle values, are in fact, compatible with the lack of Fe K$_\alpha$ line in observed spectra \citep[see e.g. ][]{2011MNRAS.415..410S}. However, higher $\beta$ values have been proposed for \bicho\ \citep{2016A&A...596A..46M}, so we let this parameter to exceed the $0.33$ limit in our fit, although, according to the reasoning presented just above, it would lead to higher $i$ values.  

The fit procedure is performed by varying each parameter value along its range in steps of 10\% of the value in the quest of the minimum reduced $\chi_\nu^2$ with $301$ degrees of freedom. Once the complete set of parameters has been fully explored, we refine the search in the vicinity of the minimum by reducing the step size. The final best-fit parameters are shown in Table \ref{model}, and the result is shown in Fig. \ref{chorros} over-plotted on the original radio maps. The northern jet is the more conspicuous and so it dominates the fit. The reduced $\chi_\nu^2 = 1.32$ is the minimum obtained when exploring a reasonable range of values in the parameter space. It is calculated based on the fit of the eight parameters and the mean error from the synthesised beam. As can be seen in Fig. \ref{chorros}, these parameters provides a reasonable match between observed jet structures and modelled paths, which suggests that studying the precession effects in the morphological evolution of \bicho\ is not out of the scope of this work. The error in each parameter is estimated with its 90\% uncertainty, obtained as the variation in this parameter which gives an increment of $e$ in the value of the global, non-reduced $\chi^2$.

\begin{table}
\caption{Twin-jet kinematic model for \object{\bicho}\ radio jets(*)}          
\label{model}      
\centering                        
\begin{tabular}{ll}     
\hline\hline                
Parameter &  Value \\  
\hline      
Angle of the precession cone             &  $\psi = (1.4 \pm 0.1)^{\circ} $    \\
Inclination of the jet precession         &  $i = (59\pm 2)^{\circ} $    \\
axis with the LOS  &  \\
Position Angle                           &  $ PA = (11.1 \pm 0.1)^{\circ} $ \\
Approaching jet (N)                      &  $s_{\rm jet} = +1 $ \\
Receding jet (S)                         &  $s_{\rm jet} = -1 $ \\
Sense of rotation (clockwise)            &  $s_{\rm rot} = -1$ \\
Precession period                        &  $P_p = (2774 \pm 11)$ d  \\
Jet velocity                             & $v_{\rm jet} = (0.330\pm0.004) c$ \\
Distance                                 & $d = (8.5\pm0.1) $ kpc \\
 \hline
 \hline                                  
\end{tabular}
~\\
(*) Adapted from \citet{1981ApJ...246L.141H}.

\end{table}

\section{Reliability of %fitted 
precession model parameters}

In order to ensure that precession is taking place on \bicho,\ more than an acceptable fitting accuracy is needed. We have to analyse the implications of the values of the parameters in Table \ref{model} and prove them to be reasonable. In this context, the distance here obtained coincides with the usually assumed value of $\sim 8.5$ kpc in the literature \citep[see e.g. ][]{1992ApJ...401L..15R,1994Meregetti,2001Keck,2001Fender,2002Grimm,2002Heindl,2005A&A...434...35H,2006Pott,2018NosotrosGRS}. However, some authors put the microquasar much closer, from $2$ kpc \citep{1992A&A...259..205M} to $5$ kpc \citep{2010Dunn}, while others place it as far as $10$ kpc based on the low radio flux and a high accretion rate observed \citep{1997Mereghetti, 1996McConnell}. We can try to constrain the distance to the source a bit more based on an independent approach founded on the relations established between extinction, $E(B-V),$ and distance, $d,$ in the Galaxy \citep[i.e.][]{1978Lucke, 1999Chen}. As $E(B-V)$ cannot be accurately measured from the overly weak companion star in \bicho,\, we can derive it from empirical relations to the column density, $N_H$ \citep{2009Guver,1995Predehl, 2006Reynoso}. As a result, assuming $N_H \simeq 1.6 \times 10^{22}$ cm$^{-2}$ from X-ray data  \citep{2011MNRAS.415..410S}, the distance to \bicho\ is $d \simeq 8-10$ kpc and $d \simeq 4-7$ kpc according to \citet{1978Lucke} and \citet{1999Chen}, respectively. Furthermore, taking into account that there is a conspicuous gas cloud possibly related to the jet-ISM interaction in \bicho\ that is located at $8.5$ kpc \citep{2018NosotrosGRS}, the distance value in Table \ref{model} seems reasonable. The existence of such a cloud near the end lobe of the jet could explain a possible slowing down of the jet velocity, lending reliability to the $\beta$ value obtained in our fitting procedure. 

On the other hand, following \citet{2011MNRAS.415..410S}, $M_\bullet \approx 5/\cos i $ M$_\odot$ for the black hole in \bicho,\, while the internal radius of the accretion disc would be $R_{in} \approx 45/\cos i $ km. Therefore, a value for the inclination, $i,$ as in Table \ref{model}, would lead to $M_\bullet \approx 10$ M$_\odot$ and  $R_{in} \approx 80$ km. It is remarkable that the mass of the black hole in \bicho\ has been derived from X-ray observations and found to be on this order \citep{2001Keck, 2006Bezayiff,  2011MNRAS.415..410S}. The internal accretion disc radius $R_{in}$ may be identified with $R_{ms}$, the radius of the marginally stable orbit of a Kerr black hole. This last parameter can be expressed in a non-dimensional form as $R_{ms}=\xi_{ms} R_g$, with $R_g = G M_\bullet / c^2$ being the gravitational radius. A dependence exists between $\xi_{ms}$ and the dimensionless spin parameter, $a,$ for the black hole such that $\xi_{ms} = 3+A_2\mp[(3-A_1)(3+A_1+2A_2)]^{1/2}$ (upper sign for prograde rotation), with $A_1 = 1+(1-a^2)^{1/3}[(1+a)^{1/3}+(1-a)^{1/3}]$ and $A_2 = (3a^2+A_1^2)^{1/2}$ \citep[see, for example,][]{1975Bardeen}. 
A radio-quiet source like \bicho\ should not have a rapidly rotating compact object \citep[at least from a theoretical point of view, see][]{1990Blandford,Tchek20} unless a geometric effect, due to the inclination of the source with respect to the observer’s line of sight combined with Doppler boosting, was at work \citep{2018motta}. However, the above observational constraints of \citet{2011MNRAS.415..410S} together with the $\xi_{ms}$ expression presented above suggest a small to moderate inclination of the \bicho\ system (always $i<60 \degree$ decreasing with the black hole spin parameter $a$). Therefore, a tentative $a = 0.1$ is  adopted from this point, which results in $R_{in} \approx 84$ km for a $10$ M$_\odot$ black hole, so the assumed value for $i$ appears to be appropriate.

However, the precession period, $P_p \sim 7.5$ years, derived from our fit is somehow troublesome. On one hand, it seems to be a reasonable value because $P_p$ is somewhere in the range of $10$--$10^2$ times the orbital period $P_o$ proposed for these kinds of sources \citep[see e.g.   ][]{1998MNRAS.299L..32L}, taking into account that the latter is accepted to be $P_o = 18.45\pm 0.10$ days from the X-ray light curve time-series analysis \citep{1538-4357-578-2-L129}. On the other hand, no long-term variability greater than $\sim 600$ days has been reported from such light curves in spite of the extensive X-ray monitoring of \bicho. In addition, a new orbital period $P_o \lesssim 1$ day has been recently proposed based on the optical observation of the companion star \citep{2016A&A...596A..46M}, which would not be compatible with $P_p \sim 7.5$ years. This new $P_o$ value has not been previously reported because it is beyond the typical Nyquist limit in X-ray lightcurves available. What is particularly notable is that this short orbital period clearly appears to be the most usual for LMXB \citep[see the BlackCAT catalogue of stellar black holes in X-ray transients,][]{2016Corral}. In addition, a significant change within the time frame of the $18.45$ day period has been reported from a detailed time-series analysis of RXTE data, ranging from $18.043$ to $18.475$ days \citep{Hirsch2020}.
This is remarkable not only because it would prevent that period from being an orbital modulation, but also because this is the observed behaviour in supra-orbital periods for other known LMXB, such as Her X-1 \citep[with 34 to 36.55 days variability, i.e.][]{2013Staubert} or SMC X-1 \citep[varying from 45 to 60 days, ][]{2019Dage}. Various mechanisms have been suggested to explain such supra-orbital modulations, but not all of them could be at work in \bicho. For instance, the magnetic warping of the disc \citep{Pfeiffer_2004} requires a highly magnetic compact object in the system, while \bicho\ appears to harbour a black-hole. A wind-driven tilting of the disc \citep{Schandl94} has been also proposed, but the plausible  A-type companion in \bicho\ is hardly capable of emitting strong enough winds. It is also difficult to relate the $\simeq 18.45$ day period to a dramatic change in X-ray state change \citep[as the origin of supra-orbital modulations in some sources, e.g.][]{King97} because they are not so dramatic nor appeared on such timescales in \bicho. More likely, the conspicuous periodicity of $\simeq 18.45$ days could be related to radiation-induced warping of the disc \citep{OD01}. However, \citet{Hirsch2020} have carefully analysed this possibility and found no definitive evidence for the conclusion that this mechanism is behind the supra-orbital motion observed in \bicho. Therefore, although we cannot strictly rule out any other mechanisms with the  data currently available, precession arises as a promising alternative to account for the (quasi-)periodic behaviour observed in X-ray light curves in \bicho.

In summary, two possible scenarios arise for \bicho. The first features an orbital period of $P_o = 18.45$ days and a precession period of $P_P \sim 7.5$ years, with radio jets approximately following the paths shown in Fig. \ref{chorros}. The most reliable companion star capable of feeding a black hole of $M_\bullet \approx 10$ M$_\odot$ through Roche lobe overflow would be a K-type giant \citep{1998Marti,2001Eikenberry}. We refer to this here as the 'K-scenario'. On the other hand, the orbital period would be $P_o \lesssim 1$ day and the precession period, $P_p$ , would be the conspicuous $18.45$ days from the X-ray light curves. In such a case, the donor star would be a main-sequence A5V \citep{2010A&A...519A..15M,2014ApJ...797L...1L, 2016A&A...596A..46M} and the system could be an intermediate-mass X-ray binary \citep{2016A&A...596A..46M,2006Tauris}. We call this the `A-scenario', which could not be related to our kinematic model fitting. Although some observational facts support of this latter case \citep{2014ApJ...797L...1L, 2016A&A...596A..46M} and it fits the $P_p/P_o$ ratio in the literature, certainly no spectroscopic feature has been obtained from the optical or IR counterpart of \bicho, so the only way to discriminate between these two pictures is to search for any lack of theoretical support.

\section{Discerning the true nature of \bicho}

For black holes, two main mechanisms have been proposed as jet drivers: the black hole spin \citep[ hereafter BZ]{1977BZ} and the disc accretion energy \citep[e.g.\,][ hereafter BP]{1982BP}. According to these two types of jet launching, for jet precession that actually takes place in \bicho,\ it would be required to have precession either of the black hole spin axis or instead the disc plane close to the central accretor. For this latter case, the inner part of the disc, where the jet
is launched, must be tilted with respect to the whole disc angular momentum. This in turn may be achieved by two mechanisms. 
%two possible physical processes have been proposed for tilting the inner part of the disc giving rise to precession. 
One possibility is to have a black hole whose spin axis is misaligned with that of the disc. In such a case, the \citet{1975Bardeen} effect can cause a local warping of the disc, forcing it to be perpendicular to the tilted black hole rotation axis. Another possibility is  based on the torques that arise via the misalignment of the accretion disc and the orbital plane of the binary \citep{1973Katz,1995Papaloizou, 1995PapaLin,1998MNRAS.299L..32L, 1997Larwood}.

Therefore, for jet precession driven by BZ, or by BP together with the Bardeen \& Petterson effect, a tilted, spinning black hole is needed. Although BZ is probably not behind the precession in our system because of the small angular momentum of the disc compared to that of the black hole \citep[which seems to be the common case, making the precession of black hole spin axes difficult to observe at human timescales; see][]{2013Nixon, 2019Banerjee}, observational evidence exists to support the idea that  
the black hole spin axis could be misaligned with the orbital and accretion disc angular momentum in X-ray binaries \citep[see e.g. ][]{HR95,2002Maccarone, Caproni2006}. This different spinning axis in accretion disc and black holes may be related to the violent birth of the system together with a corresponding long alignment timescale which has been estimated to be on the order of $\gtrsim 10^6$ years \citep{2002Maccarone}, or even higher \citep{2019Banerjee}; being, thus, greater than, or at least similar to, the system lifetime. On the other hand, jet precession may be due to a disc tilted by an external torque from the companion and a BP launching mechanisms, no matter the black hole spin. Therefore, the existence of precession in a system such as \bicho\ cannot be excluded. In the following, we try to see if this process is, in fact, taking place according to the different mechanisms mentioned above. In this sense, we ought to note that jet precession can also be driven by deflection at the collision point with super-Eddington winds coming from the disc as suggested for SS\,433 \citep{Begelman2006} or other more sophisticated mechanisms \citep{liska2018}. We cannot discuss these cases in detail for \bicho\ because of the lack of sufficiently precise observations in the vicinity of the launching site.

\subsection{Precession due to relativistic local warping}
Although the Bardeen-Petterson effect alone may be difficult to relate to precession in black hole binaries \citep{2013Nixon}, we try to study its reliability on our particular \bicho\ system. We have to consider precession and viscous timescales, $\tau_p$ and $\tau_\nu$, because the Bardeen-Petterson effect is able to align the inner part of the disc to the black hole spin axis if $\tau_p \lesssim \tau_\nu$ at the inner edge of the accretion disc $r = R_{in}$ \citep{1999Natarajan}. They are defined as:
\begin{equation}
    \tau_p = \frac{2\pi}{\Omega_p(r)} \,\,\,\,\,\,\,\,\,\,\,\,\,\,\,\, \tau_\nu = \frac{r^2}{\nu_2(r)},
\end{equation}
where $\nu_2(r)$ is the kinematic viscosity of the disc for vertical shear and $\Omega_p(r) = 2 G J_{\bullet} c^{-2}r^{-3}$ is the angular velocity due to Lense-Thirring precession \citep[e.g.][]{1972Wilkins}, with $J_{\bullet} = a G M_{\bullet}^2 c^{-1}$ as the black hole angular momentum modulus. We do not know $\nu_2(r)$ but we can assume $\nu_2 \sim \nu_1$, which is the viscosity along the disc, and can be parametrised following the \citet{1973Shakura} prescription as $\nu_1 = \alpha c_s^2 (G M_{\bullet} r^{-3})^{-1/2}$. Here, $c_s$ is the local sound speed, which in turn may be written as $c_s = H (G M_\bullet r^{-3})^{1/2}$ with $H$ being the disc scale height. So, if we associate $R_{in}$ with the $R_{ms}$ and adopt the above proposed values for \bicho\, we finally obtain:
\begin{equation}
    \frac{\tau_p}{\tau_\nu} = 10^{-1} \left(\frac{\alpha}{0.05}\right)\left(\frac{a}{0.1}\right)^{-1} \left(\frac{r}{R_{ms}}\right)^{3/2} \left(\frac{H/r}{0.1}\right)^2.
\end{equation}
Here we have assume a very conservative estimate for $H/r$ in the inner border of the disc. A reduction in one order of magnitude in this last parameter would lead to a decrease in $\tau_p/\tau_\nu$ by two orders of magnitude. Highest values of $a$ would also decrease this ratio.

\begin{figure}
   \centering
    \includegraphics[width=\hsize]{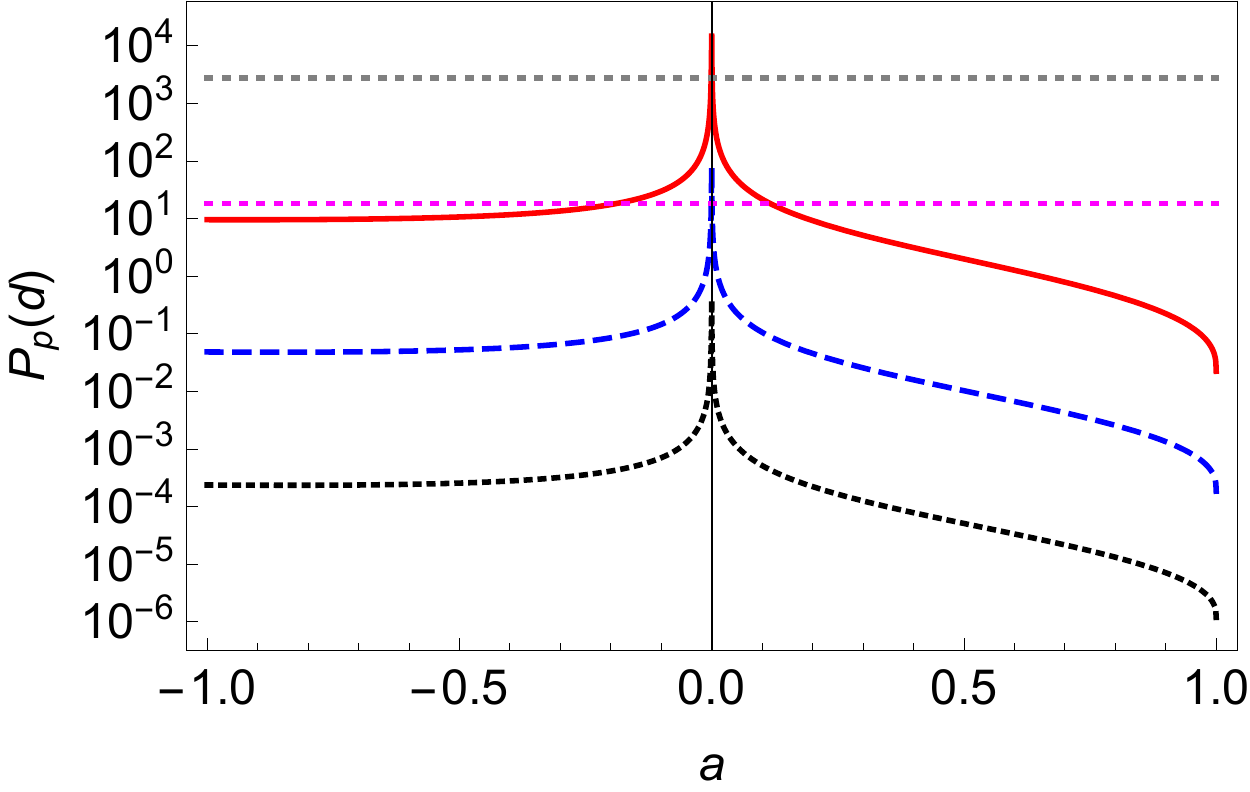}
      \caption{Precession period for \bicho\ as a function of the spin parameter. The outer non-dimensional radius of the disc is taken to be $\xi_{out}=1000 \xi_{ms}$. Different lines correspond to the different power-law slopes of the disc surface density (red-continuous for $s=0$; blue-dashed for $s=-1,$ and black-dotted for $s=-2$). The grey-dotted horizontal line marks the estimated period of precession of approximately $7.5$ yr, while the magenta-dotted line marks the $18.48$ days precession period proposed in \citet{2016A&A...596A..46M}.
      }
         \label{BP_period}
   \end{figure}

Thus, this simple estimate suggests that the precession timescale in the inner disc is likely to be shorter than the local viscous timescale, even if $H/r$ is as high as 0.3 and could be considered a thick disc which would be expected if the black hole is in the hard X-ray state. So, the precession in \bicho\ may be acting due to the Bardeen-Petterson effect even for thicker discs that could be related to jet launching at hard state. 
It is then possible to calculate the precession period of the disc or jet, taking into account the contribution of the different part of the disc using the following expression \citep{2004Caproni}:
\begin{equation}
    P_p = \frac{2 \pi G M_\bullet}{c^3}\frac{\int_{\xi_{in}}^{\xi_{out}} \Sigma(\xi)[\Phi(\xi)]^{-1} \xi^3 d\xi}{\int_{\xi_{in}}^{\xi_{out}} \Sigma(\xi) \Psi(\xi)[\Phi(\xi)]^{-2} \xi^3 d\xi},\label{pp}
\end{equation}
where $\Phi(\xi) = \xi^{3/2} + a$, $\Psi(\xi) = 1-(1-4 a \xi^{-3/2}+ 3 a^2 \xi^{-2})^{1/2}$ and $\Sigma(\xi)$ is the accretion disc surface density. Remember that $\xi$ is the  non-dimensional distance rescaled with $R_g$. We can again take $\xi_{in}=\xi_{ms}$ and assume a potential dependence of the disc density to the distance $\Sigma(\xi) = \Sigma_0\, \xi^s$ with constant parameters. We consider $s \in \{-2,-1,0\}$ to take into account reasonable decreasing or constant behaviours. As the amplitude of the precession angular velocity decreases
with the cube of the distance, the exact value of the outer radius of the accretion disc is not critical and we take $\xi_{out} = 10^3 \xi_{ms}$ as in \citet{2016A&A...596A..46M}. 
We have numerically integrated Eq. (\ref{pp}) and the resulting precession period, $P_p$, in days is plotted in Fig. \ref{BP_period} as a function of the spin parameter, $a$. The upper, grey-dotted line represents the $7.5$ years precession period of \bicho\ corresponding to the `K-scenario' defined above, while the lower, magenta-dotted line marks the precession period of $\sim 18.45$ days proposed in \citet{2016A&A...596A..46M} according to the `A-scenario'. It may be shown that the yearly precession period for \bicho\ can only be reproduced for an overly slow rotating black hole 
($a \sim 10^{-3}$) and assuming a constant density along the disc. On the other hand, the daily precession period can be reached with the here adopted rotation for the black hole ($a=0.1$) or even $a = -0.2$ for a retrograde spin under the same density conditions.  
A smaller disc will lead to shorter periods and, therefore, to more tiny, unrealistic results. If the outer disc is greater than our estimation, then precession periods grow, and a $\xi_{out}=8500$ with our assumed black hole spin $a = 0.1$ and a Sakura-Sunyaev solution $s = -\frac{3}{4}$ would result on the $18.45$ days precession period, while the $7.5$ years period is only reached if $\xi_{out} = 1.5\times10^5$ under the same conditions. These latter value is difficult to reconcile with both the observed emission from the disc and the binary separation. 
Thus, precession in \bicho, if indeed caused by the Bardeen-Petterson effect on the disc, would require an extended disc with a surface density slightly dependent on radius, and clearly would favour the `A-scenario' against the `K-scenario'.

\subsection{Precession due to external torque }
On the other hand, an external torque from the companion could maintain the disc warp and explain the precession in \bicho. We can follow \citet{1995Papaloizou} and assume rigid body precession for a differentially rotating fluid in the accretion disc, and use linear perturbation theory to derive an integral form for the precession frequency. 
Based on this, \citet{1998MNRAS.299L..32L} obtained the ratio between the binary orbital period, $P_o$, and the precession period, $P_p,$ for a non-relativistic, polytropic equation of state with ratio of specific heats of $5/3$, which is:
\begin{equation}
    \frac{P_o}{P_p} = \frac{3}{7} \frac{q}{\sqrt{1+q}} \left( \frac{R_{out}}{D}\right)^{3/2} \cos \delta,\label{pprevia}
\end{equation}
where  $\delta$ is the orbital inclination, $D$ is the orbital separation, and $R_{out}$ is the outer disc, which is estimated to be a fraction $\kappa$ of the Roche lobe around the compact object $R_{Lc}$. Here, it is assumed that $R_{out}= \kappa R_{Lc} \gg R_{in}$. This equation differs by a factor of two from the expression given in \citet{1973Katz} because the tidal torque has been averaged over the disc.  
The extension of the Roche lobe may be approximated through the expression given by \citet{1983ApJ...268..368E}:
\begin{equation}
    R_{Lc} = \frac{0.49}{0.6+q^{2/3} \ln(1+q^{-1/3})}D,
\end{equation}
so Eq. (\ref{pprevia}) may be written as:
\begin{equation}
    \frac{P_o}{P_p} = \frac{3}{7} \frac{q}{\sqrt{1+q}} \left(\frac{0.49}{0.6+q^{2/3} \ln(1+q^{-1/3}) } \right)^{3/2} \kappa^{3/2} \cos \delta. \label{popp}
\end{equation}
In this model, the inclination $\delta$ of the binary orbit with respect to the accretion disc plane gives rise to the observed precession. The jet then subtends a precession cone with a half-opening angle equal to the orbit inclination, so $\delta = \psi$, which we adopt from the
model fit ($\psi = 1.4^\circ$; see Table \ref{model}). On the other hand, the $\kappa$ value was determined by \citet{1977Pac} for a thin and weakly viscous accretion disc based on the assumption that its size could not exceed the largest orbit with non-intersecting particles in the primary Roche lobe. This parameter depends on the mass ratio, $q$, although for $0.03<q< 30$ the fraction $\kappa = 0.86$ and is roughly independent of $q$. A similar value, $\kappa = 0.88,$ was obtained from an independent approach by \citet{1977Papaprin} for $0.2<q<10$. This seems to be the right mass ratio interval for \bicho\ harbouring a $\sim 10$ M$_\odot$ black hole. 

In our `A-scenario' the $P_o/P_p = 0.03$, which is similar to the theoretical $0.02$ derived from Eq. (\ref{popp}). 
On the other hand, the `K-scenario' gives $P_o/P_p = 0.017,$ which is half the theoretical value of $0.042$ obtained from Eq. (\ref{popp}). In order to find a better discriminant argument between our two possible scenarios, we can estimate the disc size following the reasoning based on the torque on the accretion disc for a small orbital inclination binary system \citep{1998MNRAS.299L..32L}: 
\begin{equation}
    R_{out}=\left( \frac{14 \pi}{3} \frac{\sqrt{q+1}}{q}\frac{D^3}{\sqrt{G (M_\bullet + M_{donor})} P_p \cos \delta} \right)^{2/3}.\label{rout}
\end{equation}
This result is $2^{2/3}$ greater than that proposed in \citet{1997Larwood} and \citet{1995Papaloizou}, because the considered averaged effect of the donor star tidal torque on the disc. In our `K-scenario', this would result on $R_{out} \sim 10^{12}$ cm,
which is of the same order than the \citet{1977Pac} estimate of $R_{out} = \kappa R_{Lc}$. This means a non-dimensional $\xi_{out} \sim 10^5$, which is too high as said before. On the other hand, the `A-scenario' leads to a shorter $R_{out} \sim 10^{11}$ cm according both to Eq. (\ref{rout}) and Paczy\`nski estimation, which gives $\xi_{out} \sim 10^4$. We can compare these results to the $R_{out}$ estimated from the observed delay in brightening and softening events in \bicho,\ following \citet{1999Main}:
\begin{equation}
    \xi_{out} = 1 \times 10^4 \left( \frac{\alpha}{0.05}\right)^{16/25}\left( \frac{\dot{M}}{5\times 10^{17} g s^{-1}}\right)^{6/25}\left( \frac{M_\bullet}{10 M_\odot}\right)^{-1/5},\label{routrg}
\end{equation}
where $\dot{M}$ is the accretion rate, which for \bicho,\ we assume to be $\sim 5\times 10^{17}$ g s$^{-1}$ in accordance with the $\sim 0.02$ Eddington observed by \citet{2011MNRAS.415..410S} and the value adopted in \citet{1999Main}. Equation (\ref{routrg}) gives a result which is remarkably compatible with our disc size estimation in the `A-scenario'. Thus, the `K-scenario'
seems to be less realistic than the former if this precession model is at work, in addition to the yet commented disagreement to the optical and IR proposed counterpart. Moreover, an accretion disc too large should be more powerful than observed and its emission would be conspicuous at optical and IR wavelengths, in contrast to the observed spectrum \citep{2016A&A...596A..46M}. A smaller disc may be present if \bicho\ system were highly eccentric, which is not expected.

\subsection{Considering whether precession is actually taking place on \bicho}
From all the previous discussions, we can reach as a preliminary conclusion that there seems to be hints of precession in \bicho\ in any case. 
 
A precession period on the order of tens of days is favoured over one on the order of years. However, as pointed out above, such a short period would produce jet paths that are too wiggled to explain their changing morphology shown in the radio maps of Fig. \ref{chorros}, unless \bicho\ is assumed to be unexpectedly closed to us. Therefore, we propose that this source is probably configured in the `A-scenario' and that the conspicuous $\sim 18.54 $ day period has to be attributed to precession instead to orbital motion. Thus, as precession could not be responsible for the observed variations in the paths of the radio jets in \bicho, we proceed to analyse whether these could be attributed to the growth of instabilities.

\section{Instabilities as the source of radio-jet changing morphology}

Large-scale morphology changes in \bicho\ have already been attributed to the growth of instabilities \citep{2015A&A...578L..11M, 2018NosotrosGRS}, as is usual in astrophysical jets structure \citep[e.g.][]{2011Hardee, 2012Perucho}. This certainly seems to be the more promising explanation for our source, as it is evident that a lack of symmetry in the wavy patterns of the jet-counterjet structure of \bicho\ in Fig. \ref{chorros}, in contrast to confident cases of precession jets in  microquasars such as SS\,433 or 1E\,1740.7--2942. This is strongly suggestive of some kind of instability developing at large scales rather than a modulation of the jet at the central source is behind the observed evolution of the \bicho\ radio morphology.

Current-driven (CDI) is one of the possible instabilities developing in the jets if magnetic fields are present. Their azimuthal mode $m=1$ kink instability is the most effective one in modifying the magnetised jet structure \citep{2006Giannios, 2011Hardee}, leading even to its disruption. An ideal kink mode is characterised by helical displacements of the cylindrical cross sections of a plasma column. It is expected to grow on a dynamical timescale with respect to an Alfv\'en wave crossing the unstable column. 
According to simulations of AGN jets \citep[e.g.][]{2012Mizuno}, the timescale for the growth of kink instabilities is on the order of $\tau_{CDI} \sim 10 R_{jet}/v_A$, where $R_{jet}$ is the jet radius and $v_A$ the Alfv\'en speed. For \bicho,\ we measured $R_{jet} \sim 0.1$ pc \citep{2015A&A...578L..11M}, while $v_A$ could be estimated from ultrarelativistic plasma as $v_A = c B [4 \pi (u_{kin}+p_{kin})+B^2]^{-1/2}$, where $u_{kin}$ is the density of kinetic energy of the particles and $p_{kin}$ is the pressure of the ultrarelativistic plasma \citep[see e.g. ][]{1975Akhiezer}. With $p_{kin} = \frac{1}{3} u_{kin}$ for this case, and assuming equipartition between the magnetic and particle energy densities ($u_{kin} = B^2/8\pi$), we finally have $v_A = (3/5)^{1/2}c = 0.77 c$. Therefore, we find the instability growth timescale to be $\tau_{CDI}\sim 1.5 \times 10^3$ days, In Fig. \ref{chorros} we can estimate that the helical/ribbon-like pattern arises at about $\sim 50$ arcsec from the core, which for the assumed distance of $8.5$ kpc turns out to span a distance of at least $d_j \sim 2$ pc. This would imply a flow velocity $d_j/\tau_{CDI}$ greater than $c$, assuming that the instability starts close to the central engine. Therefore, the CDI kink instabilities do not appear to be responsible of the large-scale structure of \bicho.

Another possibility is the Kelvin-Helmholtz instability (KHI), which is originated from transverse velocity gradients or discontinuities due to the different kinetic energy between the jet and the typically sub-relativistic ambient medium. Ordinary modes can affect the jet surface or even the whole beam \citep[see e.g.][]{1991Birkinshaw}, leading to shocks, entrainment of ambient material, and eventually the disruption of the jet \citep{1998Bodo,2015A&A...578L..11M}. KHI seem to produce complex three dimensional helical, ribbon, and thread-like patterns \citep{2000Hardee, 2012Perucho} with an expected radio emission enhancement of about one order of magnitude. The growth timescale for KHI is of order of $\tau_{KHI} \sim (2 R_{jet}/v_j)\eta^{-1/2}$, with $\eta$ being the density contrast between the jet and the medium, and $v_j$ the jet bulk velocity \citep{2009Araudo}. 
 
Assuming $\eta \sim 10^{-1}$ for \bicho\ \citep{2015A&A...578L..11M}, and $v_j \sim c$ for establishing a lower limit, this timescale is $\tau_{KHI} \gtrsim 800$ days, which is even shorter than $\tau_{CDI}$. Therefore KHI is even less adequate to explain the surge of the helical patterns from the origin following the same reasoning as before, unless it was triggered far from the central engine because the jet would maintain its stability up to that point.

We can reach the same conclusions with an alternate reasoning as well. The helical pattern observed in the jets in the $1997$ map of \bicho\ appears to have a wavelength $\lambda \sim 15$ arcsec or $\lambda \sim 0.62$ pc at the assumed distance to the source. If the kink CDI starts near the onset of the structure, the flow speed would then be $v_j \sim \lambda/\tau_{CDI} \sim 0.5 c,$ which seems reasonable. But CDI should start close to the central engine, where the magnetic fields are strong, so it seems unrealistic for them to be responsible for the observations. However, KHI is not compelled to be triggered so near to the core. The same reasoning as before accounts $v_j \sim \lambda/\tau_{KHI} \sim c$ for the KHI starting near the onset of the wiggled structure using our $\tau_{KHI}$ lower limit, which is rather sound.   

Recently, another kind of instability that has never before considered in this context was recently proposed to explain the reconfinement and lost of stability in astrophysical jets \citep{Gourgouliatos:2018aa}. The centrifugal force acting on the plasma moving along curved streamlines promotes the so-called centrifugal instability (CFI) that induces stream-oriented features in the flow that disturbs the jet morphology. Such an instability is expected to appear if the flow satisfies the generalised Rayleigh criterion \citep{Gourgouliatos:2018aa} which, in the case of curved jets confined by an external medium at rest, is always fulfilled. 

Therefore, some instabilities seem to be capable of explaining the observed morphological changes in \bicho\ radio jets. The KHI need to start far from the central source and could be triggered by jet precession. The wiggling observed in the South jet could be reminiscent of instability-driven structures which has been found in other microquasars \citep[see e.g.  ][]{2017Migliori}. This, in addition,  strengthens the role that has been attributed to instabilities in the formation of the actual radio structures of \bicho\ \citep{2015A&A...578L..11M, 2018NosotrosGRS}.

%_____________________________________________________________

\section{Conclusions }

In this paper we have presented a study of the origin of the large scale morphological changes in the relativistic jets of \bicho. Theoretical and observational arguments led us to reach the following conclusions:

   \begin{enumerate}
      \item New radio data confirms that large-scale morphology of \bicho\ indeed changes over time in a matter of years.  The observed radio jets clearly exhibit a variable morphology with some structures resembling helical paths. 
      
      \item A kinematic model has been fit to the observed radio structures. Although some of the derived parameters may make sense, the precession period of $\sim 7.5$ years appears problematic. It would need a giant K-type companion which does not agree with previous optical/IR observations. In addition, the corresponding accretion disc would be too large for the observed emission. Therefore, the radio morphology changes in \bicho\ cannot be attributed to precession.
      
      \item However, our theoretical considerations allow for the existence of precession in \bicho. Although we cannot assure its precise origin  with the present observational evidence, actual data is compatible with precession in both Bardeen-Petterson and companion torque models. In any case, the favoured scenario is that with an A main-sequence star and an orbital period on the order of a day. This means that the precession period should be on the order of days, which is too short for being compatible with the observed path of the jets. 
      
      \item  We thus exclude precession as the source of the dramatic changes observed in \bicho. These changes have presumably been attributed to the growth of instabilities, which have shown to form ribbon-like and helical structures. The absence of symmetry in the wiggles of the jets in the large-scale structure of \bicho\ supports this view. According to our analysis, instabilities developing far away from the central source, rather than a modulation of the outflow speed or mass flux at the compact core, appear to explain the radio observations. 

      \item  As a result, we propose a new physical scenario for \bicho\ system. It could be a close binary composed of a $\sim 10$ M$_\odot$ black hole and a main-sequence A-type, $\sim 2$ M$_\odot$ donor star orbiting each other with a period $\lesssim 1$ day. Although we cannot strictly rule out other mechanisms with currently available data, as precession might certainly be at work on this system, it is suggestive to attribute to it the conspicuous observed period of $18.45$ days, which has been assumed as the orbital period up until recently. This is the only scenario compatible with available optical/IR observations and with the precession models we put into test. These properties would made \bicho\ a possible intermediate mass X-ray binary. Such a scenario would explain  its X-ray behaviour more easily, as it is more similar to high-mass X-ray binaries than to LMXB. The resulting large-scale morphology of the jets is affected by instabilities that originate the ribbon-like and helical-like structures observed. As has been previously claimed, these instabilities would eventually lead to the disruption of the jet.
   
   \end{enumerate}

\begin{acknowledgements}
      The National Radio Astronomy Observatory is a facility of the National Science Foundation operated under cooperative agreement by Associated Universities, Inc. This work was supported by the Agencia Estatal de Investigaci\'on grant PID2019-105510GB-C32/AEI/10.13039/501100011033 from the Spanish Ministerios de Ciencia e Innovaci\'on y Universidades, and by the Consejer\'ia de Econom\'ia, Innovaci\'on, Ciencia y Empleo of Junta de Andaluc\'ia under research group FQM-322, as well as FEDER funds. 
\end{acknowledgements}

%-------------------------------------------------------------------

\bibliographystyle{aa} % style aa.bst
\bibliography{references} % your references Yourfile.bib

\end{document}